\documentclass[journal,twoside, a4paper]{IEEEtran}

\usepackage{color}

\usepackage{graphicx}
\usepackage[ruled,vlined]{algorithm2e}
\usepackage[autostyle]{csquotes}
\usepackage{amsmath}

\usepackage{hyperref}
\usepackage{float}

\usepackage{comment}
\def\BibTeX{{\rm B\kern-.05em{\sc i\kern-.025em b}\kern-.08em
    T\kern-.1667em\lower.7ex\hbox{E}\kern-.125emX}}

\ifCLASSINFOpdf
\else
 
\fi

\usepackage[a4paper, total={180mm,257mm}]{geometry}

\begin{document}

\title{Exploring Image Transforms derived from Eye
Gaze Variables for Progressive Autism Diagnosis
}

\author{\IEEEauthorblockN{
Abigail Copiaco\IEEEauthorrefmark{1}, 
Christian Ritz\IEEEauthorrefmark{2}, 
Yassine Himeur\IEEEauthorrefmark{1},
Valsamma Eapen\IEEEauthorrefmark{3},
Ammar Albanna\IEEEauthorrefmark{4} and
Wathiq Mansoor\IEEEauthorrefmark{1}
}\\
\IEEEauthorblockA{\IEEEauthorrefmark{1}
College of Engineering and Information Technology, University of Dubai, UAE (acopiaco@ud.ac.ae; yhimeur@ud.ac.ae; wmansoor@ud.ac.ae)}\\

\IEEEauthorblockA{\IEEEauthorrefmark{2}School of Electrical, Computer, and Telecommunications Engineering, University of Wollongong, Australia  (critz@uow.edu.au)}\\
\IEEEauthorblockA{\IEEEauthorrefmark{3}School of Clinical Medicine
University of New South Wales, Australia (v.eapen@unsw.edu.au)
\\
\IEEEauthorblockA{\IEEEauthorrefmark{4}College of Medicine and Health Sciences, Mohammed Bin Rashid University Dubai, UAE
(albanna.md@gmail.com)}\\
}}

\maketitle

\begin{abstract}
The prevalence of Autism Spectrum Disorder (ASD) has surged rapidly over the past decade, posing significant challenges in communication, behavior, and focus for affected individuals. Current diagnostic techniques, though effective, are time-intensive, leading to high social and economic costs. This work introduces an AI-powered assistive technology designed to streamline ASD diagnosis and management, enhancing convenience for individuals with ASD and efficiency for caregivers and therapists. The system integrates transfer learning with image transforms derived from eye gaze variables to diagnose ASD. This facilitates and opens opportunities for in-home periodical diagnosis, reducing stress for individuals and caregivers, while also preserving user privacy through the use of image transforms. The accessibility of the proposed method also offers opportunities for improved communication between guardians and therapists, ensuring regular updates on progress and evolving support needs. Overall, the approach proposed in this work ensures timely, accessible diagnosis while protecting the subjects’ privacy, improving outcomes for individuals with ASD.

\end{abstract}


\IEEEpeerreviewmaketitle

\section{Introduction}

The number of individuals affected by Autism Spectrum Disorder (ASD) has increased rapidly over the past decade, with a 175\% rise in diagnosed cases during this period \cite{1}. ASD is primarily characterized by challenges in communication, attention, and behavior \cite{2}. While the severity of these challenges varies between individuals, the specialized care required often leads to substantial social and economic costs. Current diagnostic techniques, including gold standard methods such as the Autism Diagnostic Observation Schedule (ADOS) and the Autism Diagnostic Interview-Revised (ADI-R), though effective, are often time-intensive, resource-demanding, and reliant on highly trained professionals \cite{4, 5}. These limitations create bottlenecks in early diagnosis, delaying critical interventions and contributing to disparities in access to care.

Moreover, while recent advances in neuroimaging, genetics, and proteomics have uncovered potential biomarkers for ASD, their clinical applicability remains limited due to challenges such as high costs, dependence on specialized equipment, and lack of robust validation across diverse populations. Additionally, while behavioral traits such as reduced eye contact and atypical fixation patterns have been identified as significant markers for ASD \cite{3}, existing approaches to quantify and analyze these markers often rely on manual annotation or simplistic computational models that fail to fully capture the complexity of gaze dynamics.

Another significant gap lies in the lack of effective integration of image-based analysis with eye-tracking data \cite{cilia2021computer}. While eye-tracking technology has shown promise in identifying ASD-related gaze patterns, there is limited research leveraging advanced machine learning models to classify these patterns automatically \cite{colonnese2024enhancing,jaradat2024using}. Furthermore, the potential of image transformations—such as heatmaps, scan paths, and fixation maps—to enhance the representation of gaze data remains underexplored \cite{nordfalt2024utilising}. These transformations can uncover spatial, temporal, and frequency-based patterns that are crucial for accurate diagnosis but have yet to be systematically applied in ASD research \cite{xu2024autism}. Aside from this, implementing transforms to images anonymizes data and protects the privacy of the individuals concerned.

In this study, we address these gaps by proposing a transfer learning-based approach to classify visual attention patterns in children with ASD using the GoogleNet architecture. The method also evaluates Typically Developing (TD) individuals to establish a comparative baseline. Our focus is on assessing the diagnostic performance of three distinct input representations—heatmaps, scan paths, and fixation maps—derived from eye-tracking data. By analyzing key eye gaze variables, such as reduced eye contact and atypical fixation patterns, this work aims to identify early markers of ASD that are essential for timely diagnosis and intervention while protecting their privacy.  The primary contributions of this research are summarized as follows:

\begin{itemize}
\item Classify visual attention patterns in children with ASD and typically developing (TD) individuals using machine learning methods.
\item Compare the effectiveness of heat maps, scan paths, and fixation maps as input representations for detecting ASD-related visual attention anomalies, while ensuring privacy by using anonymized gaze data instead of raw facial or video recordings.
\item Identify early markers of ASD, such as reduced eye contact and atypical fixation patterns, to improve early diagnosis strategies.
\item Optimize the performance of the model through fine-tuning for each input type to achieve high classification accuracy.
\end{itemize}

\section{Background}
ASD is a complex neurodevelopmental condition characterized by challenges in social interaction, communication, and repetitive behaviors. Early and accurate diagnosis is crucial for providing timely interventions that can significantly improve outcomes \cite{hodges2020autism,new8}.

\subsection{Current Trends in Autism Diagnosis}
Traditional diagnostic frameworks, such as the Autism Diagnostic Observation Schedule (ADOS) \cite{4} and the Autism Diagnostic Interview-Revised (ADI-R) \cite{5}, remain gold standards but are resource-intensive and require specialized training. To address these limitations, researchers have proposed the use of more accessible screening tools that enable remote developmental assessments, as well as wearable devices that monitor atypical behaviors and physiological responses. These innovations provide real-time, objective data that complement traditional methods.

Firstly, genetic studies have identified numerous genes linked to ASD, while neuroimaging techniques like MRI have revealed structural and functional brain abnormalities \cite{6, 7}. Furthermore, proteomic research has identified several possible biomarkers in biological samples \cite{8}, offering promising potential for ASD diagnosis. However, despite these advancements, extensive validation across diverse populations and cases remains essential. Several studies also challenge these claims, arguing that autism is a behavioral disorder rather than a neurological condition, as current brain imaging techniques cannot reliably detect it \cite{9}.

\subsection{Artificial Intelligence for Autism Diagnosis}
Current research trends promote the potential of Artificial Intelligence (AI) in transforming autism diagnosis by enabling the analysis of large datasets and uncovering subtle patterns \cite{10, 11}. Machine learning algorithms may also be used to train on behavioral, genetic, and neuroimaging data for ASD classification. AI-based video analysis detects atypical facial expressions, gestures, and speech patterns, while natural language processing (NLP) identifies linguistic anomalies in speech development \cite{12}. Predictive models are being developed to assess ASD risk in infants and toddlers \cite{13}, facilitating earlier interventions.

Several challenges persist in the deployment of these innovations. For example, ensuring data privacy and addressing biases in datasets are critical for ethical implementation \cite{nielsen2024estimating}. Access to advanced diagnostic technologies is often limited in low-resource settings, which can lead to disparities in care \cite{bhat2024fewer}. Furthermore, many emerging tools require rigorous validation and standardization before integration into clinical practice \cite{rahman2024restoring,kim2024engagnition}.

To address these limitations, this research introduces the innovative use of eye gaze variables as a key feature for detecting autism in children. By focusing on these behavioral markers, the approach aims to provide a non-invasive and effective diagnostic tool. Additionally, the integration of image transformation techniques serves a dual purpose: enhancing data privacy by hiding sensitive information and uncovering underlying patterns and features that may remain hidden in the raw inputs. This approach not only protects user confidentiality but also enhances the model’s ability to extract meaningful insights, paving the way for more accurate and ethical autism detection techniques.

\section{Methodology}
The objective of this experiment is to classify visual attention patterns by comparing the results using three distinct types of inputs: heatmaps, scan paths, and fixation maps, as well as their image transforms. This investigation is particularly significant, as eye gaze variables, which are found to play a crucial role in detecting autism spectrum disorder (ASD) \cite{14}. The approach proposed here leverages this finding, and applies machine learning to variables derived from visual information, offering a more automated detection technique while protecting privacy.

Individuals with ASD often exhibit distinct patterns of visual attention, such as reduced eye contact, fixation on objects rather than faces, and unusual scan paths while viewing scenes \cite{14}. These eye gaze variables are critical components of the 3D and phenotypic data used in our experiment, serving as early markers of autism. Identifying these markers can facilitate timely interventions and appropriate support for individuals with ASD.

By leveraging transfer learning with GoogleNet \cite{15}, we aim to enhance the detection and classification of these visual attention patterns, thereby contributing to the early diagnosis and intervention strategies for ASD. GoogleNet offers several advantages, including its deep architecture with inception modules, which optimize computational efficiency while maintaining high accuracy \cite{15}.

\subsection{Experimental Setup}
The data was obtained from Zenodo, comprising recordings from 14 individuals in the ASD group and another 14 in the Typically Developing (TD) group, with an age range of 5 to 12 years old \cite{19}. Eye-tracking data was collected using the Tobii T120 eye tracker, featuring a resolution of 1280x1024 and a tracking distance of approximately 65 cm. Tobii Studio software was utilized to obtain eye fixation data directly \cite{20}. In total, 300 data samples were acquired for each group, for a total of 600 samples. These are categorized into three types of input data, summarized as follows:

\vspace{0.25cm}
\subsubsection{Heatmaps}
Heatmaps are visual representations of data where the values are represented by varying color intensities \cite{21}. These are typically constructed to illustrate the spatial distribution of gaze points over an image. Mathematically, heatmaps can be generated by aggregating Gaussian kernels around fixation points, which is represented in Eq.~(\ref{eq:heatmap}):
\begin{equation}
H(x,y) = \sum_{i=1}^{N} G(x - x_i, y - y_i; \sigma)
\label{eq:heatmap}
\end{equation}
where $G(x, y; \sigma)$ represents the Gaussian kernel, specified in Eq.~(\ref{eq:gaussian}), $(x_i, y_i)$ represents the fixation point coordinates, $\sigma$ represents the Gaussian spread, and $N$ is the total number of fixation points.

\begin{equation}
G(x, y; \sigma) = \frac{1}{2 \pi \sigma^2} e^{-\frac{x^2 + y^2}{2 \sigma^2}}
\label{eq:gaussian}
\end{equation}

In this application, the heatmaps were resized and normalized to match the input dimensions required by GoogleNet. The network was then trained to classify different patterns of visual attention.

\vspace{0.25cm}
\subsubsection{Scan Paths}
Scan paths represent the sequential trajectory of eye movements across a stimulus, characterized by rapid eye movements and stationary periods \cite{22}. These paths capture the temporal dynamics of gaze and provide insights into the cognitive processes underlying visual exploration. Scan paths can be described as an ordered sequence of fixations, which considers both the coordinates and duration of the fixation.

To process the scan paths as input to the model, the sequential gaze data was converted and resized into a suitable image format. The network was then trained to discern the temporal dynamics of eye movements.

\vspace{0.25cm}
\subsubsection{Fixation Maps}
Fixation maps represent the spatial distribution of stationary gaze points during visual attention \cite{23}. As opposed to heatmaps, these are usually discrete and highlight the fixation locations without any smoothing.

To process the fixation maps for GoogleNet, they were handled similarly to heatmaps and scan paths, which included resizing and normalization. The model was then trained to classify based on fixation behavior.

\begin{figure}[htbp]
\centering
\includegraphics[width=0.34\columnwidth]{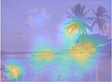}
\includegraphics[width=0.25\columnwidth]{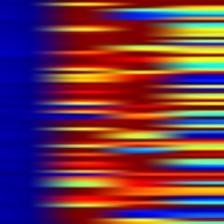}
\includegraphics[width=0.35\columnwidth]{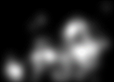}\\
~~~~~(a) Heat Map	~~~~~~(b) Scan Path	~~~(c) Fixation Map ~~~~\\
\caption{Visual Representations (a) Heat map, (b) Scan path converted into an image representation, (c) Fixation map for individuals with ASD}
\label{fig1}
\end{figure}

\begin{figure}[htbp]
\centering
\includegraphics[width=0.34\columnwidth]{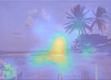}
\includegraphics[width=0.25\columnwidth]{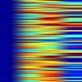}
\includegraphics[width=0.35\columnwidth]{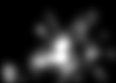}\\
~~~~~(a) Heat Map	~~~~~~(b) Scan Path	~~~(c) Fixation Map ~~~~\\
\caption{Visual Representations (a) Heat map, (b) Scan path converted into an image representation, (c) Fixation map for typically developing (TD) individuals}
\label{fig2}
\end{figure}

\vspace{0.25cm}
Fig.~\ref{fig1} displays samples of the visual data for an individual with ASD, while Fig.~\ref{fig2} displays the corresponding data for a TD individual. These images serve as input to the selected neural network model, which was trained accordingly. These images serve as input to the GoogleNet. Through observation, it can be noted that the heatmaps for children with ASD are less concentrated compared to those for TD children. Furthermore, scan paths tend to be more restricted, and fixation maps show greater clustering. This represents differences in visual attention patterns between the two groups.

Throughout the experiment, the dataset was split into 80\% for training and 20\% for testing to evaluate the model's performance. Two types of image transformations, Continuous Wavelet Transform (CWT) and Fast Fourier Transform (FFT), are utilized in this study, which are explained in the subsequent sub-sections.

\subsection{Continuous Wavelet Transform (CWT)}
The Continuous Wavelet Transform (CWT) is a type of spectro-temporal feature extraction technique that combines temporal and spectral attributes \cite{18}. This technique uses scalable wavelet functions to decompose images into a time-frequency representation. This enables the simultaneous analysis of both spatial and frequency components, making it particularly suited for capturing localized and transient features. Moreover, its ability to adapt to varying scales contributes to a robust analysis, especially in noisy environments. This ensures that intricate features are not overshadowed by more dominant patterns, enhancing its utility in complex image processing tasks.

\begin{figure}[htbp]
\centering
\includegraphics[width=0.305\columnwidth]{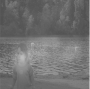}
\includegraphics[width=0.305\columnwidth]{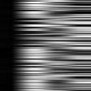}
\includegraphics[width=0.305\columnwidth]{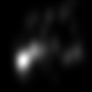}\\
~~~~~(a) Heat Map	~~~~~~(b) Scan Path	~~~(c) Fixation Map ~~~~\\
\caption{Haar Wavelet Transform Visual Representations (a) Heat map, (b) Scan path converted into an image representation, (c) Fixation map for individuals with ASD}
\label{fig3}
\end{figure}

\begin{figure}[htbp]
\centering
\includegraphics[width=0.305\columnwidth]{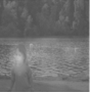}
\includegraphics[width=0.31\columnwidth]{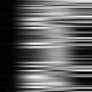}
\includegraphics[width=0.305\columnwidth]{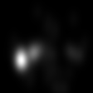}\\
~~~~~(a) Heat Map	~~~~~~(b) Scan Path	~~~(c) Fixation Map ~~~~\\
\caption{Haar Wavelet Transform Visual Representations (a) Heat map, (b) Scan path converted into an image representation, (c) Fixation map for typically developing (TD) individuals}
\label{fig4}
\end{figure}

Figs.~\ref{fig3} and ~\ref{fig4} shows examples of applying the Continuous Wavelet Transform (CWT) using the Haar wavelet as the mother filter. The Haar wavelet is advantageous due to its simplicity, fast computation, and effective time localization, making it particularly suitable for analyzing signals with abrupt changes or edges \cite{19}. These transformed images serve as inputs to the neural network model for classification tasks.

\subsection{Fast Fourier Transform (FFT)}
The Fast Fourier Transform (FFT) is an efficient algorithm that transforms signals from the time domain to the frequency domain, enabling detailed spectral analysis \cite{20}. Unlike time-domain methods, FFT emphasizes frequency components, making it highly effective for identifying periodic patterns and dominant frequencies in complex signals. When applied to images, FFT transforms spatial data into its frequency components, producing a frequency spectrum that highlights spatial frequency pattern. This differs from a spectrogram, which represents the time-frequency evolution of a signal, generated by applying a Short-Time Fourier Transform (STFT) to segmented time windows. Unlike spectrograms, which are used to analyze dynamic signals, the output of FFT on images is static, and focuses on spatial frequency distributions.

\begin{figure}[htbp]
\centering
\includegraphics[width=0.305\columnwidth]{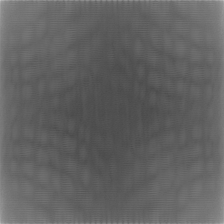}
\includegraphics[width=0.31\columnwidth]{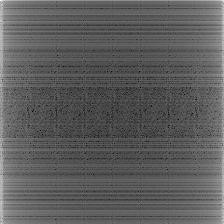}
\includegraphics[width=0.305\columnwidth]{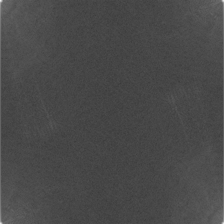}\\
~~~~~(a) Heat Map	~~~~~~(b) Scan Path	~~~(c) Fixation Map ~~~~\\
\caption{Frequency Spectrum of (a) Heat map, (b) Scan path converted into an image representation, (c) Fixation map for individuals with ASD.}
\label{fig5}
\end{figure}

\begin{figure}[htbp]
\centering
\includegraphics[width=0.305\columnwidth]{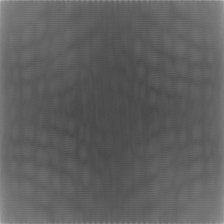}
\includegraphics[width=0.31\columnwidth]{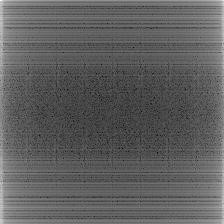}
\includegraphics[width=0.305\columnwidth]{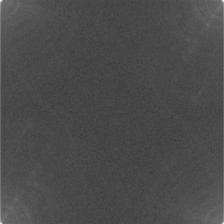}\\
~~~~~(a) Heat Map	~~~~~~(b) Scan Path	~~~(c) Fixation Map ~~~~\\
\caption{Frequency Spectrum of (a) Heat map, (b) Scan path converted into an image representation, (c) Fixation map for typically developing (TD) individuals.}
\label{fig6}
\end{figure}

Examples of the application of FFT on the original images derived from the dataset are shown in Figs.~\ref{fig5} and ~\ref{fig6}. The FFT technique is particularly valued for its computational efficiency, scalability to large datasets, and precision in spectral feature extraction. As with the CWT transform, the frequency-domain features generated by FFT are also utilized as inputs to the neural network model, enhancing its ability to classify data \cite{new2}. 

Upon closer inspection of the frequency spectrums, subtle differences can be observed in the spectral patterns between the classes. These differences are primarily reflected in the distribution of dominant frequencies and the intensity variations across the spectrum. For example, Heat maps exhibit a more concentrated set of high-frequency components, whereas Scan paths display a broader distribution with lower intensity peaks. These variations, though not immediately apparent in the figures, play a critical role in enabling the model to differentiate between classes effectively.

\subsection{Transfer Learning}
GoogleNet is a pre-trained convolutional neural network (CNN) that utilizes a Directed Acyclic Graph (DAG) architecture, enabling multiple parallel paths for feature extraction \cite{18}. This design allows layers to share outputs and inputs from multiple preceding layers, capturing a broader range of features compared to traditional sequential CNNs.

Alongside GoogleNet, AlexNet was also used in this research for comparison. AlexNet follows a simpler, sequential architecture and introduced innovations such as ReLU activation, dropout, and data augmentation \cite{alex}. Given its series format, AlexNet offers greater architectural flexibility than GoogleNet, although it is larger in size. 

Transfer learning was applied using both pre-trained models, leveraging previously learned weights to accelerate training and improve performance on the new dataset. This method significantly reduces the time and computational resources required compared to training from scratch \cite{22, 23}.

Algorithm \ref{algo1} outlines the systematic approach for ASD diagnosis using eye-tracking data and machine learning. Visual attention patterns—heatmaps, scan paths, and fixation maps—were extracted and further enhanced through transformations like Continuous Wavelet Transform (CWT) and Fast Fourier Transform (FFT). Both pre-trained models were fine-tuned using backpropagation, and performance was evaluated across various input types. Results were analyzed using accuracy metrics, training curves, and confusion matrices to identify the most effective configurations.

\begin{algorithm}[t]
\SetAlgoLined
\caption{Proposed Method for Progressive Autism Diagnosis}
\label{algo1}

\textbf{Input:} Eye-tracking data $\mathbf{D} = \{d_1, d_2, ..., d_n\}$ consisting of heatmaps, scan paths, and fixation maps for ASD and TD individuals.

\textbf{Output:} Classification results $\mathbf{y} \in \{\text{ASD}, \text{TD}\}$ with accuracy metrics.

\textbf{Steps:}
\begin{enumerate}
        \item Collect eye-tracking data using Tobii T120 eye tracker.
        \item Preprocess the data to obtain inputs $\mathbf{X}_{\text{heat}}$, $\mathbf{X}_{\text{scan}}$, $\mathbf{X}_{\text{fix}}$ corresponding to heatmaps, scan paths, and fixation maps, respectively.

        \item Apply transformations $\mathcal{T}$ to the inputs:
        \begin{equation}
            \mathbf{X}_{\text{transformed}} = \mathcal{T}(\mathbf{X}), \quad \mathcal{T} \in \{\text{CWT}, \text{FFT}\}.
        \end{equation}

        \item Load the pre-trained GoogleNet model $\mathcal{M}_{\text{pre-trained}}$.
        \item Replace the output layer to match the binary classification task: 
        \begin{equation}
            \mathcal{M}_{\text{new}} = \mathcal{M}_{\text{pre-trained}} - \text{LastLayer} + \text{BinaryOutput}.
        \end{equation}

    \ForEach{transformed input $\mathbf{X}_{\text{transformed}} \in \{\mathbf{X}_{\text{heat}}, \mathbf{X}_{\text{scan}}, \mathbf{X}_{\text{fix}}\}$}{
        \For{epoch $t = 1$ to $T$}{
                Compute predictions:
                \begin{equation}
                    \hat{\mathbf{y}} = \mathcal{M}_{\text{new}}(\mathbf{X}_{\text{transformed}}).
                \end{equation}
                 Calculate loss:
                \begin{equation}
                    \mathcal{L} = \mathcal{L}_{\text{CE}}(\hat{\mathbf{y}}, \mathbf{y}),
                \end{equation}
                where $\mathcal{L}_{\text{CE}}$ is the cross-entropy loss.
                
                 Update model weights using backpropagation:
                \begin{equation}
                    \mathbf{W}_{t+1} = \mathbf{W}_t - \eta \nabla_{\mathbf{W}} \mathcal{L},
                \end{equation}
                where $\eta$ is the learning rate.
        }
    }

        \item Evaluate the model on the test set to compute accuracy:
        \begin{equation}
            \text{Accuracy} = \frac{\text{Correct Predictions}}{\text{Total Predictions}} \times 100.
        \end{equation}

        \item Compare accuracy metrics across transformed inputs $\mathbf{X}_{\text{CWT}}$ and $\mathbf{X}_{\text{FFT}}$ for heatmaps, scan paths, and fixation maps.

        \item Visualize training curves for loss and accuracy.
        \item Display confusion matrix for the best-performing configuration.

\end{enumerate}
\end{algorithm}

\section{Results and Discussion}
\subsection{Experimental Results}
Table \ref{tab1} summarizes the results of training accuracies achieved when comparing the three inputs discussed in the previous section. These figures provide valuable insights into the performance of the neural network model across different input types and transformations, evaluated under consistent iterations and epochs.

\begin{table}[H]
\centering
\caption{Training Results Summary}
\label{tab1}
\resizebox{0.48\textwidth}{!}{%
\begin{tabular}{|l|c|c|c|}
\hline
\textbf{Input Type}           & \textbf{Iterations/Epoch} & \textbf{Epochs} & \textbf{Accuracy (\%)} \\ \hline
\multicolumn{4}{|c|}{\textbf{Pre-trained Model: GoogLeNet}} \\ \hline
Heat Maps                    & 48                       & 6               & 40.83                  \\ \hline
Scan Paths                    & 48                       & 10              & 82.50                  \\ \hline
Fixation Maps                 & 48                       & 10              & 65.00                  \\ \hline
FFT of Heat Maps              & 48                       & 10              & 45.83                  \\ \hline
FFT of Scan Paths             & 48                       & 10              & 67.50                  \\ \hline
FFT of Fixation Maps          & 48                       & 10              & 64.17                  \\ \hline
CWT of Heat Maps              & 48                       & 10              & 30.00                  \\ \hline
CWT of Scan Paths             & 48                       & 10              & 85.00                  \\ \hline
CWT of Fixation Maps          & 48                       & 10              & 65.83                  \\ \hline
\multicolumn{4}{|c|}{\textbf{Pre-trained Model: AlexNet}} \\ \hline
Heat Maps                    & 48                       & 10               & 38.73                  \\ \hline
Scan Paths                    & 48                       & 10              & 81.20                  \\ \hline
Fixation Maps                 & 48                       & 10              & 64.85                  \\ \hline
FFT of Heat Maps              & 48                       & 10              & 43.92                  \\ \hline
FFT of Scan Paths             & 48                       & 10              & 71.67                  \\ \hline
FFT of Fixation Maps          & 48                       & 10              & 68.20                  \\ \hline
CWT of Heat Maps              & 48                       & 10              & 36.67                  \\ \hline
CWT of Scan Paths             & 48                       & 10              & 79.17                  \\ \hline
CWT of Fixation Maps          & 48                       & 10              & 70.00                  \\ \hline
\end{tabular}%
}
\end{table}

\section{Analysis of Results}
As observed, among the input types, scan paths have proven to be the most effective, achieving the highest accuracy of 82.50\% without transformation and 85.00\% when combined with the Continuous Wavelet Transform (CWT). This indicates that scan paths inherently capture important sequential and spatial information, which is further enhanced by the temporal and spectral feature representation offered by CWT. On the contrary, heat maps demonstrated the least effectiveness, yielding a modest accuracy of 40.83\% without transformation and showing further declines with CWT at 30.00\%. These results suggest that heat maps do not indicate a direct relation with autism. 

For the transformed inputs, CWT proved to be a powerful transformation, particularly for scan paths, where it achieved the highest overall accuracy of 85.00\%. This highlights the advantage of CWT in capturing both temporal and spectral information, making it highly effective for tasks that require robust feature representation.

\begin{figure}[h]
    \centering
    \includegraphics[width=1\linewidth]{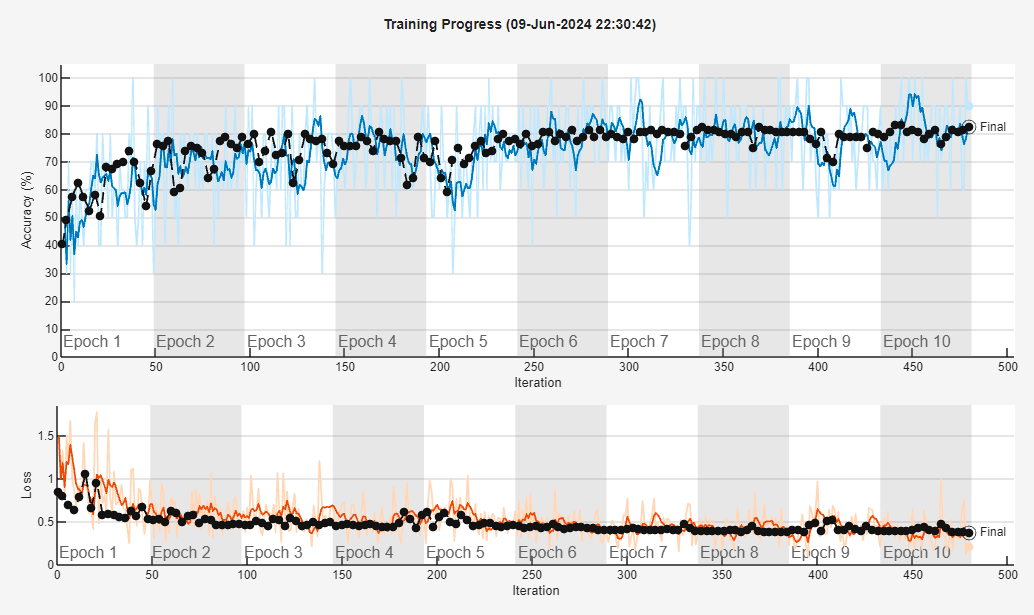} 
    \caption{Training and Loss graph for Scan Paths input to GoogleNet.}
    \label{fig:training_loss}
\end{figure}

\begin{figure}[h]
    \centering
    \includegraphics[width=0.8\linewidth]{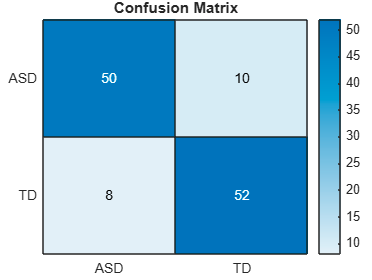} 
    \caption{Confusion Matrix – CWT of Scan Paths.}
    \label{fig:confusion_matrix}
\end{figure}

Overall, the findings highlight the critical role of the input type, as well as the image transformation or any other pre-processing techniques, in determining classification performance. Fig.~\ref{fig:training_loss} provides the training and loss graph for scan path inputs across 10 epochs at 48 iterations per epoch, while Fig.~\ref{fig:confusion_matrix} displays the confusion matrix obtained for the highest performing methodology.

\section{Conclusion}
The promising results obtained from our study open several avenues for further research. As we progress with this research, we plan on expanding the dataset to include a larger and more diverse sample size to improve the robustness and generalizability of the model. Additionally, integrating 3D data with phenotypic data, such as combining eye-tracking data with other physiological signals and ASD test scores, will provide a more comprehensive understanding of individuals with ASD.

Further exploration into advanced preprocessing techniques and data augmentation strategies for our multimodal dataset will also enhance model performance. For example, the Salient360 Toolbox presented in \cite{24} allows the visualization and comparison of eye gaze data in three-dimensional format. Provided that we confirmed the potential in improving accuracy through image transformations, refining these transformations and exploring hybrid methods to further enhance accuracy and generalizability is an interesting future direction. Furthermore, investigating alternative deep learning architectures may also yield improved accuracies by better capturing the temporal and spatial dynamics of movements.

As we progress further in this research, the real-time analysis and classification of eye-tracking data would be beneficial for the development of an interactive diagnostic and management tool, which caters to an adaptive learning environment. Finally, collaboration with clinical experts in order to validate and interpret the model's findings will also ensure that the insights gained are clinically relevant and actionable.

\section{Acknowledgement}
This work was supported by the Dubai Future Foundation under its Research, Development, and Innovation (RDI) Program. The authors thank the Foundation for its support in fostering research and innovation in the UAE.


\end{document}